# The Elastic Capacitor and its Unusual Properties

Michael B. Partensky, Department of Chemistry, Brandeis University, Waltham, MA 02453

partensky@attbi.com

The "elastic capacitor" (EC) model was first introduced in studies of lipid bilayers (the major components of biological membranes). This electro-elastic model accounts for the compression of a membrane under applied voltage V and allows one to obtain information about the membrane's elastic properties from the measurements of its capacitance. Later on, ECs were used to analyze the electrical breakdown of biological membranes. This effect is used nowadays in various medical applications, such as targeted drug delivery. The EC model was also helpful in studies of "microscopic capacitors" – electric double layers in various electrified interfaces (of which the electrode/electrolyte interface is the most common example). This comparatively simple model, the analysis of which requires only high-school physics, has a close relationship to some real-life problems in physics, chemistry and biology.

We will be examining a version of an EC model, which is a parallel-plate capacitor with an elastically suspended upper plate and a lower plate, which is fixed and grounded (Fig. 1). The separation between the uncharged plates is $h_0$, the area of each plate is $A$, and the effective "spring constant" is $KA$ ($K$ is the spring constant per unit area). As it is usually assumed, the distance $h$ between the plates is much smaller than $A^{1/2}$ and the fringe effects can be neglected. Building up charge on the plates causes the spring to stretch, until the attractive electric force and its opposing elastic force become equal, and an equilibrium position is established. The mathematical development that follows will describe the equilibrium position of the plates and the electrical parameters of the EC, and will deal with possible instabilities in the EC. The relationships that result from this seemingly simple electro-mechanical system are quite interesting, and afford one an opportunity to better understand the underlying physical processes. I hope that both teachers and students will find this discussion instructive and challenging.

## 1. Capacity anomalies in the isolated EC

### Statement of the Problem

To study the electric properties of an *isolated EC*, the charge $Q$ on the upper plate is incremented by small portions $\Delta Q$. Every time a $\Delta Q$ is added, the new value of charge stays fixed (this is what we mean by *isolated*) while $h$ and $V$ assume new equilibrium values corresponding to the new value of $Q$. _Now, students may be challenged by following questions_:

1. Describe the electro-compression in EC, i.e. $h(Q)$. At what charge $Q_{max}$ will the plates come into contact?
2. Draw the curve $V(Q)$ for $Q < Q_{max}$. Try to describe this result in terms of differential capacitance $C_d = dQ/dV \approx \Delta Q/\Delta V$. Did you find anything unusual? Try to explain the anomaly, if any.

3. How will *V(Q)* change if a stop is put in the gap at the distances $h_0/3$ or $2h_0/3$ above the lower plate, preventing the gap from further contraction?

Suggestion: try casting the problem in dimensionless units, such as $z = h/h_0$.

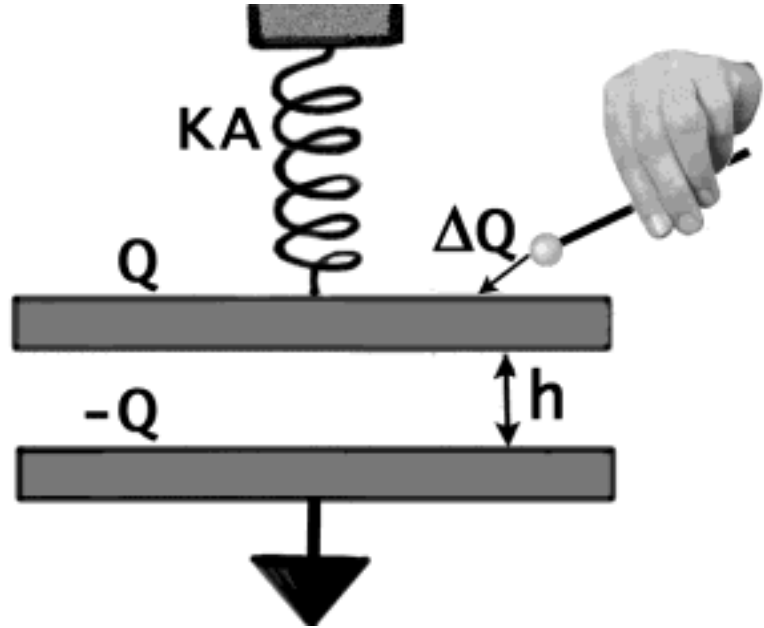

**Figure 1. Isolated Elastic Capacitor**

## **Solution**

The energy of the isolated elastic capacitor (per unit area) consists of elastic and electrostatic contributions.

$$W_{EC} = \frac{K(h-h_0)^2}{2} + \frac{\sigma^2}{2\varepsilon\varepsilon_0} h \tag{1}$$

where $\sigma = Q/A$ is the charge density and $\varepsilon$ is the dielectric constant in the gap of EC Orienting the x-axis as the upward normal to the grounded plate, we find the resultant force (per unit area) acting on the top plate,

$$F_x = K(h_0 - h) - \frac{\sigma^2}{2\varepsilon\varepsilon_0} \tag{2}$$

where the first term is the elastic force caused by the stretched spring, and the second term is the opposite force due to the electrostatic attraction of the plates. Note that the weight of the plate does not appear in Eqs. (1) and (2) because it is fixed and compensated by the elastic force. The potential difference between the plates is

$$V = \frac{\sigma}{\varepsilon\varepsilon_0} h \tag{3}$$

It is convenient to use dimensionless units,

$$\text{Energy}: \ w = \frac{W}{K h_0{}^2}; \ \text{Force}: \ f = \frac{F}{K h_0};$$

$$\text{Gap width}: z = \frac{h}{h_0}; \qquad \text{Voltage}: \ v = (\frac{\varepsilon \varepsilon_0}{K h_0{}^3})^{1/2} V; \tag{4}$$

$$\text{Charge} \quad \text{density}: \ s = (\frac{1}{\varepsilon \varepsilon_0 K h_0})^{1/2} \sigma$$

In these units, dimensionless potential, energy, and force are:

$$v = s\,z, \quad w_{EC} = \frac{(1-z)^2}{2} + \frac{s^2}{2} z, \quad f = 1 - z - \frac{s^2}{2} \tag{5}$$

The equilibrium distance between the plates is derived from the force balance condition $f = 0$:

$$z = 1 - s^2 / 2 \tag{6}$$

We now see that the gap closes ($z \to 0$) at $s \to s_{max} = \sqrt{2}$. This result answers the first question of the problem. Using equations (5) and (6), we find the potential difference as

$$v = s \cdot (1 - s^2 / 2) \tag{7}$$

The dependencies $z(s)$ and $v(s)$ are shown in Fig. 2.

At small charges ($s \leq 0.3$) EC behaves similarly to a regular (fixed) capacitor (described in Fig. 2 by $v_0(s)$): the potential is practically a linear function of the charge. However, as $s$ grows, $v(s)$ progressively flattens, and it finally reaches its maximum at

$$s_{cr} = \sqrt{\frac{2}{3}} \approx 0.81, \quad z_{cr} = \frac{2}{3}, \qquad v_{cr} = \frac{2}{3}\sqrt{\frac{2}{3}} \approx 0.54 \tag{8}.$$

Students can verify these "critical" values either numerically (using Eq. 7), or by finding the maximum from the differential condition $dv / ds = 0$. For $s > s_{cr}$, $v(s)$ becomes the descending function of charge. In other words, *an increase of electric charge on the plates is accompanied by a decrease of the potential drop across the capacitor* [1]

---

[1] To better understand this unusual behavior, a kinematical analogy may be useful. As you know, an automobile moving with constant velocity v covers a distance L= v t in time t. Suppose that you drive a "super car" that can run as fast as you wish, and your final goal is to cover a maximum distance. However, in order to make your life more difficult, a restriction is imposed. The higher the velocity you chose, the shorter becomes the time you are allowed to move. In fact, the time is explicitly described as a descending function of v: t = $t_0$ (1 - 0.5 $v^2$), where $t_0$ is a given constants. Try to find the optimal velocity v=$v_{max}$ by drawing the function L= v t(v) and finding at what v=$v_{max}$ it reaches maximum. You will discover that $v_{max}$= $\sqrt{1/3}$. After this problem is solved, note that it exactly relates to the EC problem through the substitution v → s, t(v) → z(s): the reduction of time depending on $v$ is analogous to the contraction of the capacitor gap depending on charge.

This unusual feature becomes even more remarkable when described in terms of (dimensionless) differential capacitance [2] per unit area, $C_d \approx \Delta s / \Delta v$ where $\Delta s$ is a small variation of charge density and $\Delta v$ is a corresponding variation of potential. Those familiar with derivatives can use a precise definition of $C_d$: $C_d = ds / dv = (dv / ds)^{-1} = [1 - (s / s_{cr})^2]^{-1}$. As $s$ approaches $s_{cr}$, $C_d \to \infty$. Note, that in a fixed capacitor $C_d$ becomes infinite only in the limit $z \to 0$. In contrary, in an elastic capacitor at the point $s = s_{cr}$ where $C_d$ becomes infinite, gap width $z$ is finite ($z = 2 / 3$). The most intriguing consequence of our equations, however, is that $C_d(s)$ becomes negative for $s > s_{cr}$.

This unusual feature also appears in studies of various elaborate models of the "microscopic capacitors" (electrical double layers). As in ECs, it is generally caused by some sort of electro-compression, although the role of a spring is usually played by a combination of molecular, electrostatic and "entropic" forces [1]. The discussion of these effects and reality behind them might be a topic for a student's physics project.

## Suggestions for further study

**1.** We did not answer the question about the influence of a stop inserted in the gap of the EC. This question might be offered for independent study.

**2.** Our analysis was based on the equilibrium condition $f = 0$. It is always a good idea to verify if a discovered equilibrium is stable. This is equivalent to the requirement that the upper plate resides in a minimum of the energy curve $W(z)$, not in a maximum. If the contrary were true, then all our previous results including the anomalies of $C_d$ would become invalid. The students can be asked to depict the dependencies $w_{EC}$ for at least three characteristic values of charge, $s < s_{cr}$, $s = s_{cr}$ and $s > s_{cr}$ corresponding respectively to the regions with $C_d > 0$, $C_d = \infty$ and $C_d < 0$. They will find that each curve has a single minimum, which means that the equilibrium is stable.

**3.** The EC model that we used does not cover one very important feature of membranes or double layers. In reality, their "plates" are not rigid. For example, a membrane's flexibility allows a lateral variation of $h$ and a corresponding local variation of charge density $\sigma$ (needed to maintain the "conductive plates" as equipotentials). How can such flexibility alter our results? Once again, the EC can be helpful in answering this question [1].

The simplest example, is to use two identical ECs connected in parallel. It turns out that in a range of charges where the capacitance of a "rigid plate" EC would become negative, the "flexible plate" EC (modeled by two ECs in parallel) loses its stability. Such instabilities and "phase transitions" in membranes and charged interfaces are well known, very important and widely studied.

---

[2] When the properties of a capacitor depend on its charge (or Voltage), the differential capacitance becomes a far more appropriate tool than the regular (integral) capacitance, $C = s / v$ (total charge by total Voltage). In a sense, using $dv / ds$ instead of $v/s$ is similar to using the instantaneous ($v$) rather than the average ($\bar{v}$) velocity to describe an accelerated motion. Obviously, instantaneous ("differential") characteristics provide much more complete and unambiguous information about the system than its averaged properties. For instance, knowing that $\bar{v}=0$ one would still not be able to distinguish between a resting sate and a state of accelerated motion with $v$ changing its direction.

**4.** The charge-controlled conditions discussed in this chapter are very rare in the studies of real microscopic capacitors, such as electrical double layers and membranes. It is much more practical to connect the electrodes of a measuring cell to a potential source (battery) so that voltage $V$ (not the charge) is under control. The students could be challenged to study the EC's properties in such an "extended" system. Particularly, they can be asked to analyze how the EC's stability depends on $V$. These questions are discussed below. See also Ref. [2], the problem 25-62.

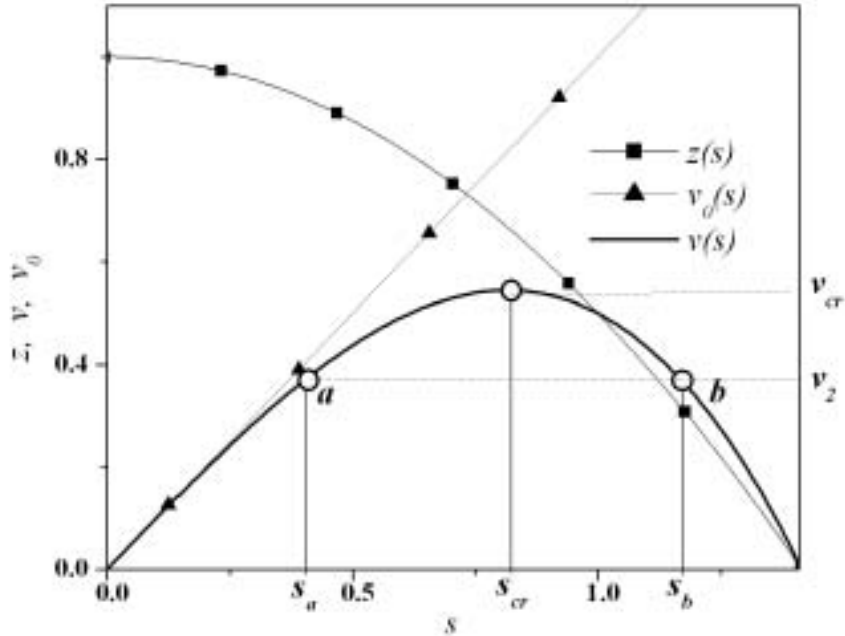

**Figure 2.** Dependencies of the gap width $z$, voltage $v$ and its components, $v_0$ and $v_1$ (with the opposite sign) on the charge density $s$ (dimensionless units). The meaning of the points $a$ and $b$ will be explained in Section 2.

## 2. The Elastic Capacitor under Potential Control

We now consider the EC connected to a battery, when the potential difference $V$ is controlled. In such an "extended" system, charge can be exchanged between the EC and the battery. Using equations (2) and (3), the electric force. $F_x^{electr}$ between the plates can be expressed in terms of $V$ as $F_x^{electr} = -V^2 / 2\varepsilon\varepsilon_0 h^2$ where the minus indicates that the force is attractive. The charging energy for EC connected to the battery is $W_{ext}^{electr} = -V^2 / 2\varepsilon\varepsilon_0 h$. To study the properties of the extended system, we use the dimensionless units of Eq. 4. The last two equations in (5) should be replaced respectively by the equations

$$w_{ext} = (1-z)^2/2 - v^2/2z \qquad (9)$$

and

$$f = 1 - z - v^2/2z^2 \qquad (10)$$

It is worth noting that the equilibrium distance derived from the condition $f = 0$ can still be expressed through the charge density using Eq. 6. Only now, instead of being an independent variable, $s$ represents the equilibrium charge density corresponding to a fixed potential $v$, and must be expressed through $v$ using Eq. 7: $s(1 - s^2/2) = v$. The equilibrium values of $s$ can be found graphically as the $s$-coordinates of the intersection points between a horizontal $v = const$ and the curve $v(s)$ (Eq. 7). Fig. 2 shows that there are two distinct regions of potential:

(a) for any $v < v_{cr}$ there are two equilibrium solutions, such as the points **a** and **b** shown for $v = v_2$ (the meaning of $v_2$ will be clear from Fig. 3) .

(b) for $v > v_{cr}$ no such intersections exist, and the extended system does not have any equilibrium state. The critical potential, $v_{cr}$, separates these two regions. Now it is appropriate to ask which of the two equilibrium solutions, $s_a$ or $s_b$, should be chosen. In other words, which one corresponds to a stable equilibrium? To answer this question we can study the profiles $w_{ext}(s)$ for different values of $v$, preliminary making in Eq. 10 a substitution $z \to v/s$ (see Eq. 4). Several such profiles are shown in Fig, 3. In the range $0 < v < v_{cr}$ each curve has a well (minimum) and a hump (maximum). As the potential increases, the minimum shifts towards a larger $s$, while the hump shifts in the opposite direction. The second curve in Fig. 3 corresponds to $v = v_2$ of Fig 2. Comparing these two figures we find that the smaller of two equilibrium charge densities shown in Fig. 2, $s_a$, corresponds to minimum of the energy, while $s_b$ corresponds to maximum This finding actually answers our question: the equilibrium corresponding to the solution $s = s_b$ (Fig. 2) is unstable. In other words, the range of charges that corresponds to the equilibrium of the expanded system is $s < s_{cr}$. In this range of charge $C$ is positive.

Therefore we should conclude that the differential capacitance under $V$-control is strictly positive. This result, which we obtained specifically for the EC model, is in fact universal and applicable to all sorts of capacitors, both macroscopic and microscopic.

At critical potential (curve 3) the well and the hump merge, and the equilibrium disappears. As a result, the elastic capacitor collapses. This phenomenon is closely related to the electric breakdown in membranes. Let us now estimate a breakdown voltage for a typical lipid membrane. The equilibrium thickness of the lipid bilayer (the basic component of the membranes) is $h0 \sim 2.5$ nm, its effective elastic constant K$\sim$10-11 N/nm, and dielectric constant $\varepsilon \sim 2$ .

Using Eqs. 4 and 8 we find that $V_{cr} = 0.54 \sqrt{Kh_0^3/\varepsilon\varepsilon_0} \approx 2$ V . This value correctly represents the order of magnitude of the membrane breakdown voltage, although the experimental values are typically 2-3 times lower. To explain those differences would require a much more elaborate model of the membrane which is beyond the scope of this paper [1].

Finally, it should be noticed that the electric breakdown described above is similar, but not identical to a phenomenon widely know as "dielectric breakdown" [2], where a material loses its insulating properties when a sufficiently strong electric field (exceeding the "dielectric strength" of the material) is applied [3].

---

[3] It is interesting to notice that some microscopic capacitors can hold electric fields far exceeding the typical "dielectric strength" values for insulators. For example, the charge density in contacts of metal electrodes with electrolytes (liquid or solid) can reach $10 - 20 \ \mu C/cm^2$, which corresponds to electric field strength $F \sim 10^9 \ V/m$ existing at microscopic distances ($\sim 1 \ nm$) near the interface. Note that the highest dielectric strengths of macroscopic materials can hardly reach 1/10 of this value.

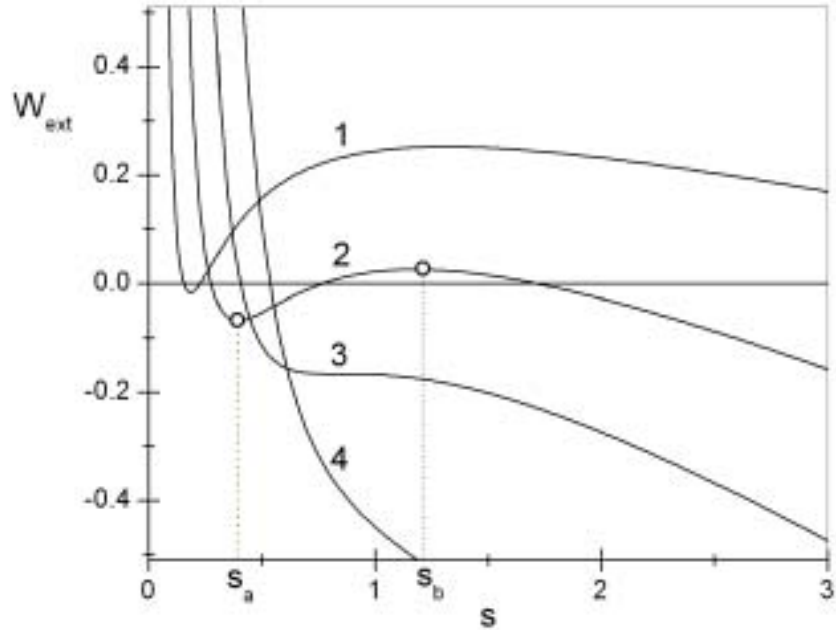

**Fig. 3:** Energy profiles *w(z)* for the expanded system for different values of the potential. The curve numbered *n* corresponds to the potential $v_n = n\, v_{cr}/3$.

## *Acknowledgement*

I am grateful to Vitaly J. Feldman and Peter C. Jordan for their valued insights, and to John Griffin, Barry Cohen and Joseph Cox for helpful comments.